\documentclass[12pt]{article}
\usepackage{epsf}
\usepackage{amsmath,amssymb}
\usepackage{slashed}
\usepackage{graphicx}

\allowdisplaybreaks

\usepackage{comment}
\usepackage{cite}
\usepackage{hyperref}
\usepackage{cases}

\begin{document}

\newcommand{\rem}[1]{{$\spadesuit$\bf #1$\spadesuit$}}

\renewcommand{\thefootnote}{\fnsymbol{footnote}}
\setcounter{footnote}{0}

\begin{titlepage}

\def\thefootnote{\fnsymbol{footnote}}

\begin{center}

\hfill UT-19-26\\
\hfill November, 2019\\

\vskip .75in

{\Large \bf

  Signals of Axion Like Dark Matter\\
  in Time Dependent Polarization of Light\\

}

\vskip .5in

{\large
  So Chigusa, Takeo Moroi and Kazunori Nakayama
}

\vskip 0.5in

{\em
Department of Physics, University of Tokyo, Tokyo 113-0033, Japan}

\end{center}
\vskip .5in

\begin{abstract}

  We consider the search for axion-like particles (ALPs) by using time
series data of the polarization angle of the light.  If the
condensation of an ALP plays the role of dark matter, the polarization
plane of the light oscillates as a function of time and we may be able
to detect the signal of the ALP by continuously observing the
polarization.  In particular, we discuss that the analysis of the
Fourier-transformed data of the time-dependent polarization angle is
powerful to find the signal of the ALP dark matter.  We pay particular
attention to the light coming from astrophysical sources such as
protoplanetary disks, supernova remnants, the foreground emission of
the cosmic microwave background, and so on.  We show that, for the ALP
mass of $\sim 10^{-22}$--$10^{-19}\ {\rm eV}$, ALP searches in the
Fourier space may reach the parameter region which is unexplored by
other searches yet.

\end{abstract}

\end{titlepage}

\renewcommand{\thepage}{\arabic{page}}
\setcounter{page}{1}
\renewcommand{\thefootnote}{\#\arabic{footnote}}
\setcounter{footnote}{0}
\renewcommand{\theequation}{\thesection.\arabic{equation}}

\section{Introduction}
\label{sec:intro}
\setcounter{equation}{0}

The existence of axion-like particles (ALPs) may be ubiquitous in
string theory~\cite{Svrcek:2006yi, Arvanitaki:2009fg, Cicoli:2012sz}
and they may have a wide range of masses and decay constants. In
particular, a very light ALP is a candidate for dark matter (DM) in the
present universe.  The ALP field, denoted by $a$, begins coherent
oscillation when the Hubble parameter becomes comparable to the ALP
mass and it behaves as a non-relativistic matter.  Thus finding the
evidence of such ALP DM would be a probe of physics beyond the
Standard Model.  Actually, several ideas are proposed to search for ALP
DM through the (extremely) weak interaction between the ALP and the
Standard Model particles~\cite{Sikivie:1983ip, Horns:2012jf,
  Budker:2013hfa, Sikivie:2013laa, Rybka:2014cya, Sikivie:2014lha,
  Kahn:2016aff, Barbieri:2016vwg, DeRocco:2018jwe}.

One important effect of ALP is on the polarization plane of the light
propagating through the ALP condensation \cite{Harari:1992ea}.  If the
ALP amplitude depends on time, which is the case for the ALP DM, the
polarization plane of the light becomes also time-dependent.  Thus, if
the ALP plays the role of DM, we have a chance to observe the effects
of ALP condensation by precisely observing the polarization of the
light.  In particular, if we consider the light from astrophysical
sources, which travels a significant amount of distance before being
observed, the effects of the ALP condensation may be accumulated in
the polarization plane of the light; such an effect may be
experimentally detectable.  The ALP search using the polarization of
light from astrophysical sources have been considered in literatures,
using the light from radio galaxies~\cite{Alighieri:2010eu},
protoplanetary disks~\cite{Fujita:2018zaj}, jets in active
galaxies~\cite{Ivanov:2018byi},
pulsars~\cite{Liu:2019brz,Caputo:2019tms}, and the cosmic microwave
background (CMB)~\cite{Lue:1998mq,Finelli:2008jv,
  Pospelov:2008gg,Zhao:2014yna,Galaverni:2014gca, Fedderke:2019ajk}.
Even if the ALP is not DM, the axion cloud may be
formed around rotating black holes through the
superradiance~\cite{Brito:2015oca}, and the effect on the polarization
of light passing through such axion cloud was discussed in
\cite{Plascencia:2017kca,Chen:2019fsq}.  In particular, in
\cite{Ivanov:2018byi,Liu:2019brz,Caputo:2019tms,
  Finelli:2008jv,Fedderke:2019ajk,Plascencia:2017kca,Chen:2019fsq},
possibilities of using the time dependence of the polarization of
light were discussed.

In this letter, we consider how we can extract information about the
ALP DM from the time dependence of the polarization of the light from
astrophysical sources.  Assuming that the ALP is the dominant
component of cold DM, the ALP potential should be well approximated by
a parabolic one in the present universe.  In such a case, the
time dependence of the polarization plane becomes also
harmonic-oscillator-like with the angular frequency of $m_a$ (with
$m_a$ being the mass of the ALP).  With such knowledge about the
time dependence of the polarization plane, we can extract information
about the ALP from the behavior of the polarization plane by Fourier
transforming the time-dependent data.  While the ALP search in the
Fourier space has already been done by using jets from active galaxies
\cite{Ivanov:2018byi} and radio pulsar \cite{Caputo:2019tms}, we
discuss that we may use polarized light from a variety of
astrophysical light sources for the ALP search, like polarized light
from protoplanetary disk, astrophysical radio sources like supernova
remnant (SNR), foreground emission of the CMB, and so on.  We show
that the analysis with the Fourier transformation applies to these light sources.  We estimate the possible discovery
reaches for the ALPs using these astrophysical polarized light; our
formulation is generally applicable to various sources and can be used
as a guideline for future ALP search.  We will argue that, with the
use of the information about the time dependence of the polarization
plane of the light, the discovery reach for the ALPs can be
significantly enlarged.  We also show that the ALP search of our
proposal may explore the parameter region which has not been reached
by the experiments in the past.

The organization of this letter is as follows.  In Section
\ref{sec:methodology}, we give the setup of our analysis and explain
how the information about the ALP DM can be extracted from the data.
In Section \ref{sec:sensitivity}, we estimate the sensitivity of the
ALP search by observing the time dependence of the polarization plane
of astrophysical light.  We give a general formula which gives the
reach to the strength of the ALP-photon coupling as a function of the
ALP mass, observation time, angular resolution for the measurement of
the polarization plane, and so on.  Then, we estimate the expected
sensitivities of ALP search using lights from various sources, such as
protoplanetary disks, radio sources like SNRs, CMB, and so on.
Section \ref{sec:conclusions} is devoted to conclusions and
discussion.

\section{Setup and Methodology}
\label{sec:methodology}
\setcounter{equation}{0}

We first explain the methodology of how we can perform the ALP search
by observing the polarization of lights from astrophysical sources.
We adopt the following Lagrangian:
\begin{align}
  {\cal L} =
  -\frac{1}{4} F_{\mu\nu} F^{\mu\nu}
  + \frac{1}{2} \partial_\mu a \partial^\mu a
  - \frac{1}{2} m_a^2 a^2
  + \frac{1}{8} g \epsilon^{\mu\nu\rho\sigma} a F_{\mu\nu} F_{\rho\sigma},
\end{align}
where $F_{\mu\nu}\equiv\partial_\mu A_\nu-\partial_\nu A_\mu$ is the
field strength for the electromagnetic gauge boson $A_\mu$, $a$ is the
ALP field, and $g$ is the photon-ALP coupling constant.  (Here, we
neglect terms irrelevant for our discussion.)

We assume that the universe is filled with the ALP oscillation that is
assumed to be DM.  Because the interaction of the ALP is extremely
weak, it obeys the equation of motion of free bosons.  We approximate
the ALP amplitude at the position $\vec{x}$ as
\begin{align}
  a_{\vec{x}} (t) \simeq
  \tilde{a}_{\vec{x}}
  \sin
  \left( m_a t + \delta_{\vec{x}} \right).
  \label{a(x,t)}
\end{align}
We will come back to the validity of the above expression later.  The
amplitude of the oscillation is related to the (local) DM
density as
\begin{align}
  \rho^{\rm (ALP)}_{\vec{x}} =
  \frac{1}{2} m_a^2 \tilde{a}_{\vec{x}}^2.
\end{align}
We can find
\begin{align}
  \tilde{a}_{\vec{x}} \simeq 2.1 \times 10^9\ {\rm GeV}
  \times
  \left( \frac{m_a}{10^{-21}\ {\rm eV}} \right)^{-1}
  \left( \frac{\rho^{\rm (ALP)}_{\vec{x}}}{0.3\ {\rm GeV/cm^3}} \right)^{1/2}.
\label{amplitude}
\end{align}

Now we discuss how the light propagates under the influence of the ALP
oscillation.  The equation of motion of $A_\mu$ is given by
\begin{align}
  \Box A^\mu -
  \partial^\mu \partial_\nu A^\nu
  + g \epsilon^{\mu\nu\rho\sigma}
  (\partial_\nu a) (\partial_\rho A_\sigma) = 0.
  \label{EOM}
\end{align}
Because we are interested in the polarization of the light, we
concentrate on the plane-wave light and pay particular attention to
the electric field $\vec{E}$ perpendicular to the direction of the
propagation.  We take $z$ (i.e, $x^3$) direction as the propagation
direction, and consider the behaviors of $E_{x} = \partial_t A_{x} -
\partial_{x} A_0$ and $E_{y} = \partial_t A_{y} - \partial_{y} A_0$.

Hereafter, we take the Lorentz gauge,
\begin{align}
  \partial_\mu A^\mu = 0,
\end{align}
and discuss how the solutions of Eq.\ \eqref{EOM} behave.  Because the
effect of the ALP is so weak that it can be treated as a perturbation,
we discuss the effect of the ALP on $A_\mu$ up to the linear order in
$g$.  Then, $A_\mu$ can be expressed as
\begin{align}
  A_\mu = (0, A_x, A_y, 0) + \delta A_\mu,
\end{align}
with $\delta A_\mu\sim O(g)$, where we have used the fact that, in the
vacuum, we can take a gauge in which both $A_0=0$ and $\partial_iA_i=0$ hold.
Besides, we concentrate on the case where
\begin{align}
  \omega_\gamma \gg m_a,
\end{align}
with $\omega_\gamma$ being the angular frequency of the light of our
interest.

For our discussion, it is convenient to use the fact that the
following equation holds in the Lorentz gauge (with neglecting terms
of $O(g^2)$ or higher):
\begin{align}
  \Box (A_x + i A_y) = - i g
  \left[ (\partial_t a) \partial_z - (\partial_z a) \partial_t \right]
  (A_x + i A_y).
  \label{EoM}
\end{align}
Because $\omega_\gamma \gg m_a$, $\partial_t a$ and $\partial_z a$ can
be approximately treated as constants in solving Eq.\ \eqref{EoM}.
Postulating
\begin{align}
  A_x + i A_y =
  \tilde{A}_-
  \exp \left[ - i \{ \omega_\gamma t - k_- (\omega_\gamma) z \} \right]
  + \tilde{A}_+
  \exp \left[  i\{ \omega_\gamma t - k_+ (\omega_\gamma) z \} \right],
\end{align}
with $\tilde{A}_\pm$ being constants, we can find the following
dispersion relation:
\begin{align}
  k_\pm (\omega_\gamma) = \omega_\gamma \mp
  \frac{1}{2} g \frac{d a}{d \ell},
\end{align}
where
\begin{align}
  \frac{d a}{d \ell} \equiv
  \partial_t a + \partial_z a.
\end{align}
Then, integrating over the path of the light, parameterized as
$(t_{\rm S}+\ell,0,0,z_{\rm S}+\ell)$, we obtain
\begin{align}
  (A_x + i A_y) (t, \vec{x}) =
  \tilde{A}_-
  e^{-i\Phi_- (t, \vec{x}; t_{\rm S}, \vec{x}_{\rm S})}
  + \tilde{A}_+
  e^{i\Phi_+ (t, \vec{x}; t_{\rm S}, \vec{x}_{\rm S})}
\end{align}
with
\begin{align}
  \Phi_\pm (t, \vec{x}; t_{\rm S}, \vec{x}_{\rm S}) =
  \pm \frac{1}{2}
  g \left[ a (t, \vec{x}) - a (t_{\rm S}, \vec{x}_{\rm S}) \right],
\end{align}
where $t_{\rm S}$ is the time when the light is emitted while
$\vec{x}_{\rm S}$ is the position of the source.  Thus, $A_\mu$ is affected by the
change of the phase velocity of the light; we can
see $\delta A_{x,y} \sim O(gaA_{x,y})$.

We can also estimate the size of $A_0$.  In the Lorentz gauge, $A_0$
obeys
\begin{align}
  \Box A_0 =
  - g \epsilon^{0ijk}
  (\partial_i a) (\partial_j A_k)
  \simeq \pm i \omega_\gamma g \epsilon^{0i3k} (\partial_i a) A_k,
\end{align}
where, in the second equality, we used the fact that $A_k$ is
approximately proportional to $e^{\pm i\omega_\gamma(t-z)}$.  Notably,
$A_0=0$ if the ALP field is homogeneous (in other words, $v\rightarrow
0$ with $v$ being the velocity of the ALP).  Because of the oscillatory behavior of the ALP, we can postulate
as $A_0\propto e^{\pm i(\omega_\gamma\pm m_a)t \mp i\omega_\gamma z}$
and estimate the size of $A_0$.  Using the relation $\partial_i a\sim
O(m_a v a)$, $A_0$ is found to be of $O(gavA_\pm)$ and hence, compared
to $\delta A_\pm$, $A_0$ is suppressed by the factor of $\sim v$.
Thus, we can neglect the effects of $A_0$ on $E_x$ and $E_y$ as far as
the ALP is non-relativistic.

If the light is linearly polarized when it is emitted, the
polarization plane becomes tilted after the propagation by the angle
\begin{align}
  \Phi = \Phi_+ - \Phi_-
  =
  g \left[  a (t, \vec{x}) - a (t_{\rm S}, \vec{x}_{\rm S}) \right].
  \label{TiltAngle}
\end{align}
\begin{itemize}
\item If the source is small enough, we can use the above expression.
  Then, with Eq.\ \eqref{a(x,t)}, the polarization plane we may
  observe behaves as
  \begin{align}
    \Phi_\odot (t)
    =
    g \left[
      \tilde{a}_\odot \sin (m_a t + \delta_\odot)
      - \tilde{a}_{\rm S}
      \sin \left\{ m_a (t-L) + \delta_{\rm S} \right\}
      \right]
    ~~:~~
    \text{small source},
    \label{Tilt-Solar}
  \end{align}
  where $L$ is the distance to the source.  (Here and hereafter, the
  subscript ``$\odot$'' is used for quantities in the solar system
  unless otherwise mentioned.)  We emphasize that the polarization
  plane oscillates with the angular frequency of $m_a$.
\item If the size of the source (which we call $\Delta L$) is
  non-negligible, lights with various $L$ contribute to the signal.
  In such a case, the second term of the right-hand side of Eq.\
  \eqref{TiltAngle} should be replaced by the average, $\langle a
  (t_{\rm S}, \vec{x}_{\rm S})\rangle_L$.  In particular, if $\Delta
  L\gg m_a^{-1}$, we expect $\langle a (t_{\rm S}, \vec{x}_{\rm
    S})\rangle_L$ to vanish, and
  \begin{align}
    \Phi_\odot (t)
    =
    g \tilde{a}_\odot \sin (m_a t + \delta_\odot)
    ~~:~~
    \langle a(t_{\rm S}, \vec{x}_{\rm S})\rangle_L \sim 0.
    \label{eq:large_source}
  \end{align}
  Notice that the above expression is also applicable to the case when
  the DM density at the source is much smaller than that at
  the observer.
\end{itemize}

We consider the situation in which we measure the polarization of the
light coming from astrophysical sources.  We assume that an area of
the sky can be observed or scanned within the time $\tau$.  The
observation is assumed to be continuously performed during $0<t<T$.
The observed area is divided into smaller independent patches; the
number of sky patches is denoted as $n_{\rm patch}$, and the direction
to the patch $\alpha$ is indicated as $\vec{e}_\alpha$.  Then, each
sky patch is observed $n_{\rm obs}$ times, where
\begin{align}
  n_{\rm obs} =\frac{T}{\tau},
\end{align}
and each observation time is represented by
\begin{align}
  t_I \equiv \tau I ~~~(I=1, 2, \cdots, n_{\rm obs}).
\end{align}

For each patch $\alpha$, we assume that the direction of the
polarization on the plane perpendicular to $\vec{e}_\alpha$ is
observed as a function of time; we denote the observed value of the
polarization angle as $\Theta_\alpha(t)$.  We define the angle
relative to the time-averaged value as
\begin{align}
  \theta_\alpha (t_I) \equiv \Theta_\alpha (t_I)
  - \frac{1}{n_{\rm obs}} \sum_{J=1}^{n_{\rm obs}} \Theta_\alpha (t_J).
\end{align}
It is expected to be expressed as
\begin{align}
  \theta_\alpha (t_I) =
  \left[
    \Phi_\odot (t_I)
    - \frac{1}{n_{\rm obs}} \sum_{J=1}^{n_{\rm obs}} \Phi_\odot (t_J)
  \right]
  + N_\alpha (t_I),
  \label{theta}
\end{align}
where $N_\alpha (t_I)$ is the effect unrelated to the ALP oscillation
(which we call ``noise''), like the detector noise and observational
errors.  We also define $N_\alpha (t_I)$ such that its averaged value
vanishes (i.e., $\lim_{T\rightarrow\infty}\frac{\tau}{T}\sum_IN_\alpha
(t_I)=0$).  The variance of $N_\alpha$ is given by
\begin{align}
  \sigma_N^2 \equiv
  \frac{1}{n_{\rm obs}} \sum_{I=1}^{n_{\rm obs}} N_\alpha^2 (t_I).
  \label{sigma_N}
\end{align}
Here and hereafter, we assume that (i) the variance does not depend on
the patch, (ii) there is no correlation between $N_\alpha (t_I)$ and
$N_\beta (t_J)$ if $\alpha\neq\beta$ or $t_I\neq t_J$, and (iii)
$N_\alpha$ is uncorrelated with $\Phi_\alpha$.  In Eq.\ \eqref{theta},
the effect of the time dependence of the source, which may add an extra
contribution to $\theta_\alpha$, is not included.  We assume that such
an effect is negligible and study the reach for the case that the
sensitivity is noise limited.  The effect of the time dependence of
the source should strongly depend on the choice of the source, and its
detailed study is beyond the scope of this letter.  We expect,
however, that the effect of the time dependence of the source is
distinguishable from the signal unless it is characterized by a
frequency with a very narrow bandwidth.  This is because, as we see
below, the signal is given by a very sharp peak in the Fourier space.

For the time interval $T$ shorter than $\sim 1/(m_av^2)$,
$\theta_\alpha (t_I)$ approximately behaves as
\begin{align}
  \theta_\alpha (t_I) = A_* \sin (m_a t_I + \delta_*) + N_\alpha (t_I)
  - \frac{A_*}{n_{\rm obs}} \sum_{J=1}^{n_{\rm obs}}
  \sin (m_a t_J + \delta_*),
  \label{theta_I}
\end{align}
where $A_*$ and $\delta_*$ are constants. Taking $\tilde{a}_{\rm
  S}=\tilde{a}_\odot$ and $\delta_{\rm S}=\delta_\odot$ in
Eq.\ \eqref{Tilt-Solar}, for example,
$A_*=2g\tilde{a}_\odot\sin\frac{m_aL}{2}$ and
$\delta_*=\delta_\odot-\frac{1}{2}m_a L+\frac{1}{2}\pi$.

To obtain information about the ALP oscillation, we introduce the
following quantity:
\begin{align}
  S (\omega, \delta) \equiv
  \sum_{\alpha=1}^{n_{\rm patch}}
  \sum_{I=1}^{n_{\rm obs}} \theta_\alpha (t_I) \sin (\omega t_I + \delta).
  \label{fnS}
\end{align}
We require $\omega$ to satisfy
\begin{align}
  \frac{2\pi}{T} \lesssim \omega \lesssim \frac{2\pi}{\tau},
  \label{eq:req1}
\end{align}
with which the constant term in the right-hand side of Eq.\
\eqref{theta_I} can be made unimportant in the calculation of $S
(\omega, \delta)$.  Then, we obtain
\begin{align}
  S (\omega, \delta)
  = &\,
  \frac{1}{2} n_{\rm patch} A_*
  \sum_{I=1}^{n_{\rm obs}}
  \left[
    \cos \left\{ (\omega -m_a) t_I + \delta - \delta_* \right\}
    -
    \cos \left\{ (\omega +m_a) t_I + \delta + \delta_* \right\}
    \right]
  \nonumber \\ &\,
  + \sum_{\alpha=1}^{n_{\rm patch}}  \sum_{I=1}^{n_{\rm obs}}
  N_\alpha (t_I) \sin (\omega t_I + \delta).
\end{align}
When $\tau\ll m_a^{-1}$, the first summation can be approximately
replaced by the integration over time, i.e.,
\begin{align}
  \sum_{I=1}^{n_{\rm obs}}
  \cos \left\{ (\omega \pm m_a) t_I + \delta \pm \delta_* \right\}
  \simeq &\,
  \frac{\sin \{ (\omega \pm m_a) T\}}{(\omega \pm m_a)T}
  n_{\rm obs} \cos (\delta \pm \delta_*)
  \nonumber \\ &\,
  - \frac{1 - \cos \{ (\omega \pm m_a) T\}}{(\omega \pm m_a)T}
  n_{\rm obs} \sin (\delta \pm \delta_*).
\end{align}
The function $S(\omega,\delta)$ fluctuates as we vary $\omega$ because
of the noise.  Using the fact that $N_\alpha (t_I) \sin(\omega
t_I+\delta)$ ($I=1$, $\cdots$, $n_{\rm obs}$) are random variables
with the variance of $\frac{1}{2}\sigma_N^2$, the typical size of $S$
is estimated as
\begin{align}
  S (\omega, \delta) \sim
  \left\{
  \begin{array}{ll}
    \displaystyle{
      \frac{n_{\rm tot}}{2}A_* \cos (\delta - \delta_*)
      + \sqrt{\frac{n_{\rm tot}}{2}} \sigma_N
    } &
    ~:~ \displaystyle{
      | \omega - m_a | \ll \frac{1}{T}
    },
    \\[4mm]
    \displaystyle{
      \sqrt{\frac{n_{\rm tot}}{2}} \sigma_N
    }&
    ~:~ \displaystyle{
      | \omega - m_a | \gg \frac{1}{T}
    },
  \end{array}
  \right.
  \label{S(statav)}
\end{align}
where
\begin{align}
  n_{\rm tot} = n_{\rm obs} n_{\rm patch}.
\end{align}
With choosing $\delta\sim\delta_*$, the function $S (\omega, \delta)$
has a peak at $\omega= m_a$ when $\sqrt{\frac{n_{\rm tot}}{2}} |A_*|$ is
substantially larger than $\sigma_N$.  Thus, with a large enough
number of data, we may be able to observe the effect of the ALP even
if $\sigma_N\gg |A_*|$.

\begin{figure}[t]
  \centering
  \includegraphics[width=0.7\linewidth,bb=0 0 562 420]{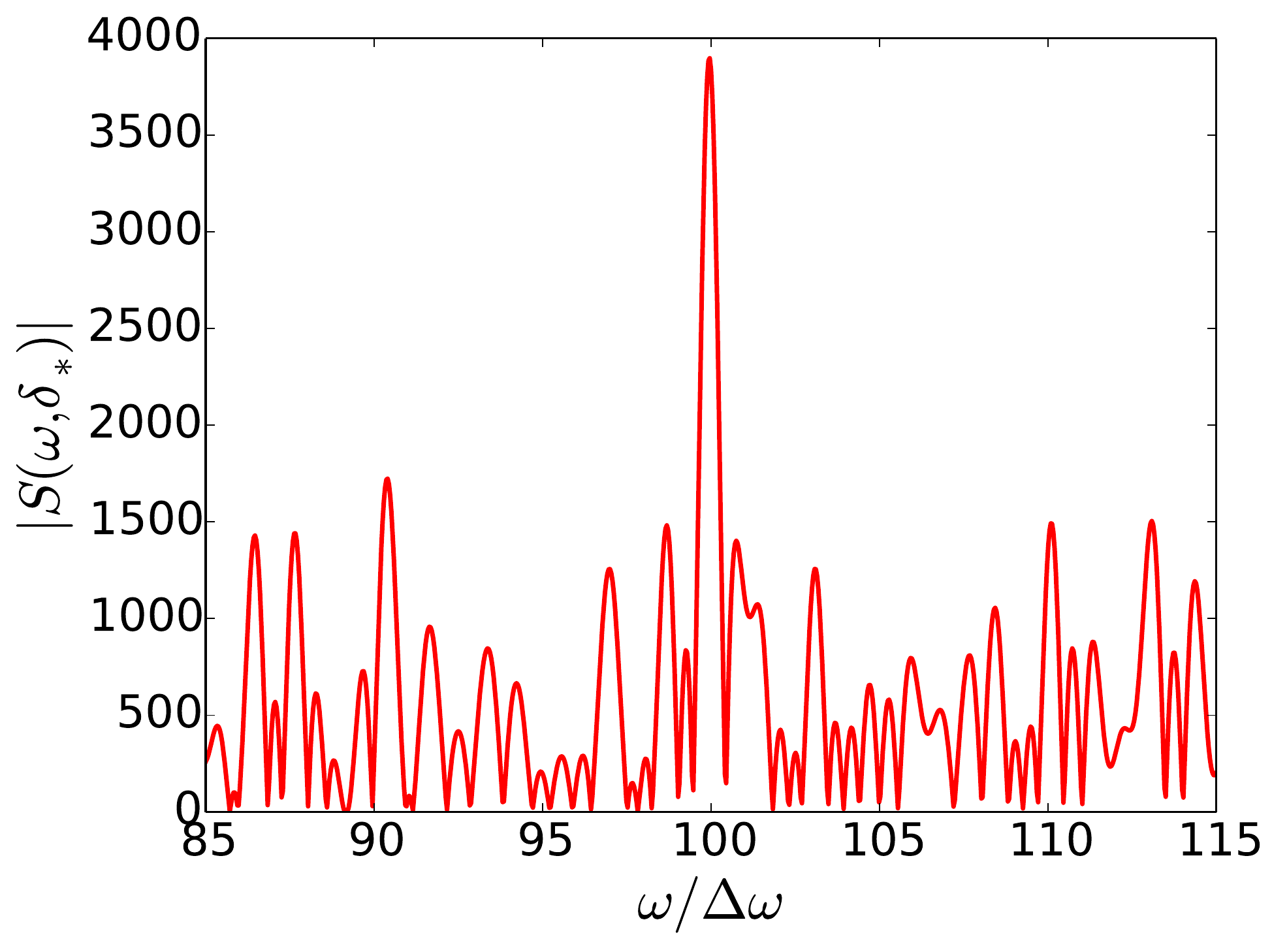}
  \caption{Behavior of $|S(\omega,\delta_*)|$ based on MC data
    generated with $\sigma_N=1$, $n_{\rm obs}=n_{\rm patch}=1000$,
    $A_*=5\sqrt{2/n_{\rm tot}}$, and $m_a=100\Delta\omega$.}
  \label{fig:signalfn}
\end{figure}

For a demonstration purpose, we generate a set of Monte-Carlo (MC)
sample data of $\theta_\alpha (t_I)$, taking
$\sigma_N=1$,
$n_{\rm obs}=n_{\rm
  patch}=1000$, $A_*=5\sqrt{2/n_{\rm tot}}$, and
$m_a=100\Delta\omega$ with
\begin{align}
  \Delta\omega = \frac{2\pi}{T}.
\end{align}
Here, the noise $N_\alpha (t_I)$ is assumed to be $(0,1)$
Gaussian random variable.  Then, we calculate the signal function $S$
with $\delta=\delta_*$; the behavior of $|S(\omega,\delta_*)|$ around
$\omega\sim m_a$ is shown in Fig.\ \ref{fig:signalfn}.  We can see
that $|S|$ is sharply peaked at $\omega= m_a$; the width of the peak
is $\sim \Delta\omega/2$.

In order to understand how large $T$ can be, we consider the validity
of Eq.\ \eqref{a(x,t)}.  Because the ALP evolves under the influence
of the gravitational potential and also because its amplitude has
fluctuations, the ALP field is given by the superposition of the modes
with various frequencies.  We expect that the ALP amplitude is {\it
  locally} expressed as
\begin{align}
  a_{\vec{x}} (t) =
  \int d\omega_a\, \tilde{a}_{\vec{x}} (\omega_a)
  \sin
  \left[ \omega_a t + \delta_{\vec{x}} (\omega_a) \right].
\end{align}
We also expect that the ALP oscillation is dominated by the modes with
$\omega_a-m_a\lesssim O(m_av_{\vec{x}}^2)$, where $v_{\vec{x}}$ is the
infall velocity, and hence $\tilde{a}_{\vec{x}} (\omega_a\gg
m_a+m_av_{\vec{x}}^2)\sim 0$.  (Near the solar system, the infall
velocity is $v_\odot\sim 10^{-3}$.) Thus, for the time period shorter
than $\sim 1/(m_av_{\vec{x}}^2)$, we may use Eq.\ \eqref{a(x,t)}; for
$|t|\lesssim 1/(m_av_{\vec{x}}^2)$, $\tilde{a}_{\vec{x}}$ and
$\delta_{\vec{x}}$ in Eq.\ \eqref{a(x,t)} are given by
\begin{align}
  \tilde{a}_{\vec{x}} \simeq
  \sqrt{s_{\vec{x}}^2 + c_{\vec{x}}^2},~~~
  \delta_{\vec{x}}  \simeq
  \cos^{-1} \frac{c_{\vec{x}}}{\sqrt{s_{\vec{x}}^2 + c_{\vec{x}}^2}},
\end{align}
where
\begin{align}
  s_{\vec{x}} = \int d\omega_a\, \tilde{a}_{\vec{x}} (\omega_a)
  \sin \delta_{\vec{x}} (\omega_a),~~~
  c_{\vec{x}} = \int d\omega_a\, \tilde{a}_{\vec{x}} (\omega_a)
  \cos \delta_{\vec{x}} (\omega_a).
\end{align}
For the time period longer than $\sim 1/(m_av_{\vec{x}}^2)$, on the
contrary, the coherence of the modes with different frequencies breaks
down; in such a case, Eq.\ \eqref{a(x,t)} is not applicable.  Thus,
our estimation of $S (\omega, \delta)$ given in Eq.\ \eqref{S(statav)}
is valid only when $T\lesssim 1/(m_av_{\vec{x}}^2)$.

\section{Sensitivity}
\label{sec:sensitivity}
\setcounter{equation}{0}

We have seen that the signal of the ALP may be imprinted in the
polarization of the lights from astrophysical sources, and that it may
be extracted if the information about the time dependence of the
polarization angle is available.  Hereafter, we estimate the
sensitivity of the ALP search using various polarized lights from
various sources.

For $m_a\gtrsim T^{-1}$, the sensitivity using the polarized light can
be obtained by solving $\sqrt{\frac{n_{\rm tot}}{2}} |A_*| \gtrsim
Z\sigma_N$, where $Z$ is the significance.  Taking $|A_*| \sim
g\tilde{a}_\odot$, the reach for the case of $m_a\gtrsim T^{-1}$ is
found to be
\begin{align}
  g \gtrsim
  2.1\times 10^{-15}\ {\rm GeV}^{-1}
  \times Z
  \left( \frac{m_a}{10^{-21}\ {\rm eV}} \right)
  \left( \frac{T}{1\ {\rm year}} \right)^{-1/2}
  \left( \frac{\tau / n_{\rm patch}}{1\ {\rm sec}} \right)^{1/2}
  \left( \frac{\sigma_N}{1^\circ} \right),
  \label{eq:sensitivity}
\end{align}
where Eq.\ \eqref{amplitude} is used to estimate the ALP amplitude
with taking $\rho^{\rm (ALP)}_\odot=0.3\ {\rm GeV/cm^3}$.  For
$m_a\lesssim T^{-1}$, on the contrary, the observation period $T$ is
too short to observe the oscillatory behavior of the signal and hence
we expect no sensitivity.
Note that $\tau$ cannot be made arbitrarily small since the error $\sigma_N$ may increase for smaller $\tau$.
There is an optimal choice of $\tau$ in a realistic experimental setup.

\begin{figure}[t]
  \centering
  \includegraphics[width=0.7\linewidth,bb=0 0 560 410]{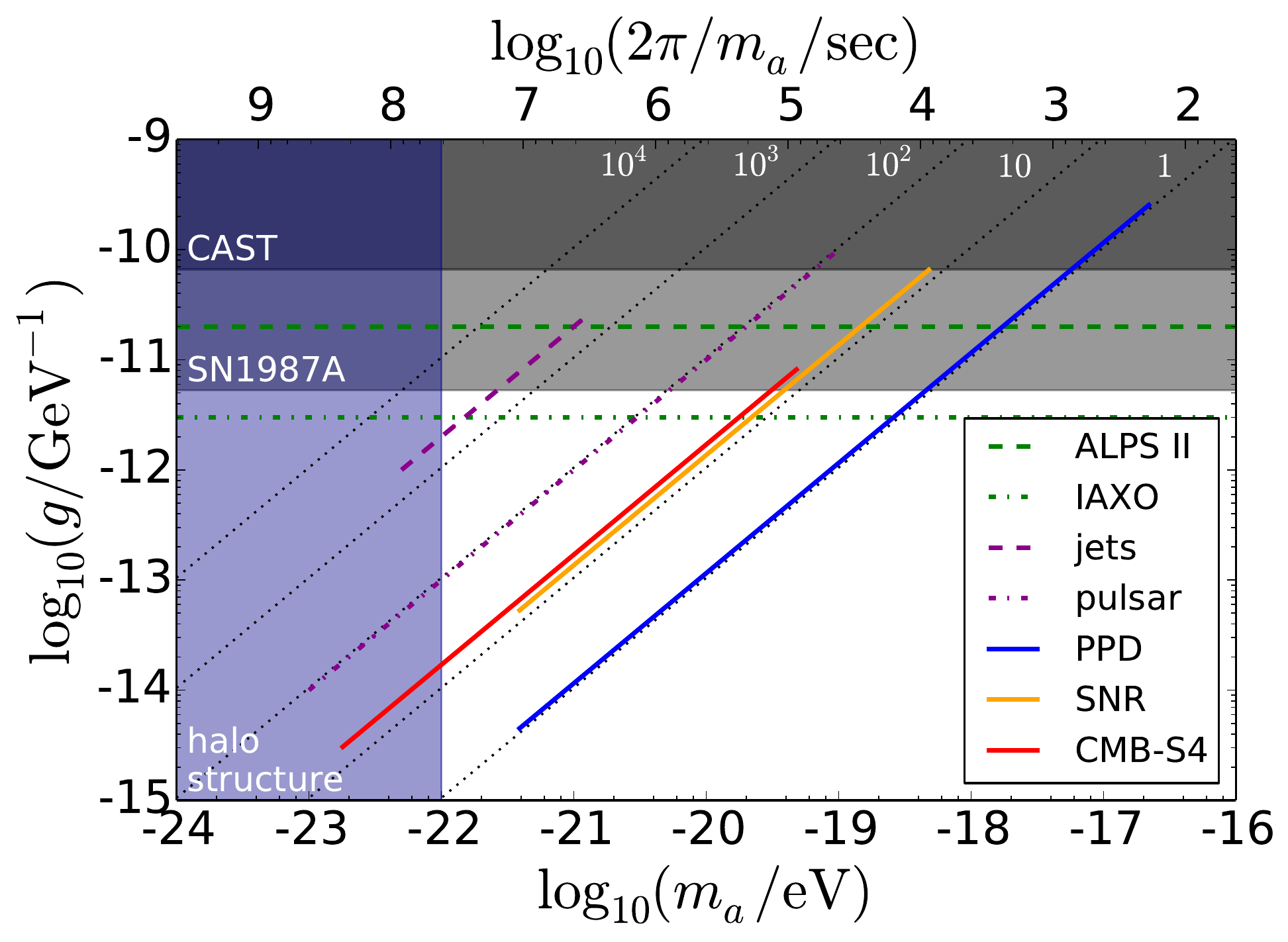}
  \caption{Expected $5\sigma$ discovery reaches on $m_a$ vs.\ $g$
    plane; the regions above the lines will be explored with the
    observations of the time dependence of the polarization plane of
    astrophysical light.  The black dotted lines are for $(T/{1\ {\rm
        year})^{-1/2}(\tau/n_{\rm patch}/1\ {\rm sec}})^{1/2}
    (\sigma_N/1^\circ)=1$ to $10^4$.  The solid lines show the
    expected $5\sigma$ reaches using a PPD with $T=4\, {\rm months}$
    (blue), an SNR with $T=4\, {\rm months}$ (orange), and by the
    CMB-S4 with $T=7\, {\rm years}$ (red), with our choices of
    detector parameters.  The end points of these lines are determined
    by the requirement \eqref{eq:req1}.  The current and the future
    constraints on the plane are also shown.  The gray shaded regions
    are existing bounds from the CAST experiment
    \cite{Anastassopoulos:2017ftl} and the observation of the SN1987A
    \cite{Payez:2014xsa}, while the blue shaded region are the
    theoretical bound from the galaxy-scale structure
    \cite{Hu:2000ke}.  The green lines are the prospects by the ALPS
    II \cite{Ehret:2010mh, Bahre:2013ywa} and IAXO
    \cite{Armengaud:2014gea} experiments.  The purple lines indicate
    bounds from the observation of parsec-scale jets
    \cite{Ivanov:2018byi} and the pulsar PSR J$0437$-$4715$
    \cite{Caputo:2019tms}.  }
  \label{fig:sensitivity}
\end{figure}

In Fig.\ \ref{fig:sensitivity}, we show the $5\sigma$ discovery reach
calculated with Eq.~\eqref{eq:sensitivity} by black dotted lines,
taking $(T/{1\ {\rm year})^{-1/2}(\tau/n_{\rm patch}/1\ {\rm
    sec}})^{1/2} (\sigma_N/1^\circ)=1$ -- $10^4$.  In the same figure,
we also show the region already excluded by the CERN Axion Solar
Telescope (CAST) experiment \cite{Anastassopoulos:2017ftl} and the
astronomical observation of the supernova SN1987A
\cite{Payez:2014xsa}, as well as by the theoretical requirement for
the galaxy-scale structure \cite{Hu:2000ke}.  Furthermore, the green
lines show the prospects of the experimental reaches by the Any Light
Particle Search (ALPS) II \cite{Ehret:2010mh, Bahre:2013ywa} and
International Axion Observatory (IAXO) \cite{Armengaud:2014gea}.
Although not shown in the Figure, the X-ray observation of sources in or behind galaxy clusters
also give stringent upper bound~\cite{Wouters:2013hua,Berg:2016ese,Marsh:2017yvc,Conlon:2017qcw}.

In the following, we discuss several examples of astrophysical sources
of polarized light for the ALP search.  In particular, we estimate the
reach for each example.

\subsection*{Protoplanetary disks (PPDs)}

PPDs are often formed around young stars.  The light emitted from the
young star and scattered in the PPD can be highly polarized, which may
be used for the ALP search of our proposal.  (For another attempt of
the ALP search using a PPD without relying on the time dependence, see
\cite{Fujita:2018zaj}.)

One of the precise measurements of the polarized intensity (PI) of PPD
can be found in \cite{Hashimoto:2011xe} for the Herbig Ae star AB Aur,
which is about $144\ {\rm pc}$ away from the earth.  In
\cite{Hashimoto:2011xe}, linear polarization images of AB Aur were
obtained by using Subaru / HiCIAO; the field of view
was $10''\times 20''$ with the pixel scale of $9.3\ {\rm mas/pixel}$,
and the total integration time of the PI image was $189.6\ {\rm sec}$.
The field of view was divided into the ``sky patches'' of the size of
$6\times 6$ pixel binning, and the polarization vector of each sky
patch was determined; the direction of the polarization vector of each
sky patch was defined as the relative angle between the polarization
vector and the line from AB Aur to the sky patch.  Using $\sim 1400$
sky patches, each of which has the PI larger than $50\sigma$, the
distribution of the polarization direction was obtained.  With the
Gaussian fitting, the distribution of the direction of the
polarization vector has the central position of $90.1^\circ\pm
0.2^\circ$ and its FWHM is $4.3^\circ\pm0.4^\circ$.

Based on the result of \cite{Hashimoto:2011xe} as well as those of
other observations of PPDs \cite{Lucas:2004jv, Tamura:2006vh,
  Hashimoto:2012, Uyama:2018}, we expect that precise measurements of
the PI of PPDs are possible.  Adopting the performance given in
\cite{Hashimoto:2011xe} as an example, let us estimate the sensitivity
of the ALP search with the use of polarized light from PPDs.  For this
purpose, we first estimate the size of the noise, $\sigma_N$.  Here,
conservatively, we assume that the FWHM of the distribution of the
polarization angle obtained in \cite{Hashimoto:2011xe} is fully due to
the effects of the noise, although the polarization angle may receive
other effects; we use $\sigma_N\simeq 1.8^\circ$ for the measurements
of the polarization angle of the light from PPDs.  Taking $\tau=
189.6\ {\rm sec}$ and $n_{\rm patch}=1400$, a 4-month observation of a
PPD similar to AB Aur gives a $5\sigma$ discovery of the ALP signal if
$g \gtrsim 1.1 \times 10^{-14}\ {\rm GeV}^{-1}\times (m_a/10^{-21}\
{\rm eV})$.\footnote
{The reach here is obtained with the assumption of 24-hour
  observation of the PPD each day.  Such an assumption may be
  unrealistic in particular for observation in the optical band that is
  possible only at night.  If the daily observation time is reduced
  down to 8 hours, for example, the reach is worsened by the factor of
  $\sqrt{3}$.}
The $5\sigma$ reach for the setup described above is shown in
Fig.~\ref{fig:sensitivity} by the blue line.  According to the
requirements Eq.~\eqref{eq:req1}, the line extends within the
range\footnote
{The upper bound on $m_a$ given in \eqref{eq:blue} is comparable to
  that from the coherent time.}
\begin{align}
  \frac{2\pi}{T} \sim 4 \times 10^{-22}\, \mathrm{eV}
  \lesssim m_a \lesssim
  \frac{2\pi}{\tau} \sim 2 \times 10^{-17}\, \mathrm{eV}.
  \label{eq:blue}
\end{align}
We can see that, for the ALP mass $m_a \lesssim
10^{-19}\,\mathrm{eV}$, we may reach a parameter region which is not
constrained by any existing bound.

\subsection*{Radio sources}

Synchrotron emission from the electron motion in the magnetic field is
necessarily polarized.  There are many astrophysical sources of such
polarized emission: radio galaxies, pulsar, SNR, pulsar wind nebula,
and so on.  In the radio frequency band, the linearly polarized light
usually experience birefringence while it is propagating through the
intergalactic plasma and magnetic field (i.e., Faraday effect). It can
be distinguished from the ALP-induced birefringence by using the
multi-wavelength observation since the former depends on the
wavelength while the latter does not.  Since we are interested in the
periodic change of the polarization angle, the Faraday effect does not
much affect our argument.

Crab nebula is one of the bright SNRs seen in a broad frequency range
from radio to gamma-ray.  The polarization angle and its distribution is
measured in optical band~\cite{Moran:2013cla} and also in radio
band~\cite{Aumont:2009dx,Ritacco:2018dsu}.  For example, the NIKA
instrument at the IRAM telescope~\cite{Ritacco:2018dsu} measured the
polarization spatial distribution of the Crab nebula at the 150\,GHz
frequency. With the observation time of 2.4 hours, the number of patches
is about $n_{\rm patch}\sim 150$ with an angular resolution of $18''$ and
the typical error of the polarization angle at each patch is about
$1^\circ$.
In Fig.~\ref{fig:sensitivity}, we show the corresponding bound with an orange line,
assuming the same total observation time as the PPD observation mentioned above.
Compared with the case of PPD observation by the Subaru / HiCIAO,
an expected constraint on $g$ is an order of magnitude weaker. Still we can
get a meaningful constraint on $g$ around the mass range $m_a \sim
10^{-21}$--$10^{-20}$\,eV.

Other polarized radio sources include jets in active
galaxies~\cite{Ivanov:2018byi} and pulsars~\cite{Liu:2019brz,Caputo:2019tms} as we
mentioned in Section\ \ref{sec:intro}.  They already have time series
observational data of the polarization angle and derived constraint on
$g$.  The observational data of the pulsar PSR
J0437-4715~\cite{Yan:2011bq} was used in \cite{Caputo:2019tms}. During
the whole 1609 days of observation, there were 393 data points of the
polarization angle with a typical variance of a few degrees, which
resulted in the constraint $g\lesssim 10^{-12}\,{\rm GeV^{-1}}\times
(m_a/10^{-21}\,{\rm eV})$.  The constraint is roughly consistent with
Eq.~(\ref{eq:sensitivity}).

\subsection*{Galactic center}

Polarized light from the galactic center may be also used for the ALP
search. In particular, linearly polarized radio waves are observed
from Sgr A*, where a supermassive black hole (BH) is expected to
exist.  The BH is associated with magnetic fields, which cause the
synchrotron radiation of the polarized light.  The BH mass is measured
to be $M_{\mathrm{BH}} \sim 4.1 \times 10^6 M_{\odot}$ (with $M_\odot$
being the solar mass), while the position of the radiation source is
$\sim 10 R_S$ away from the center of the BH \cite{Bower:2018wsw},
with $R_S \sim 10^{-6} \mathrm{pc}$ being the Schwarzchild radius.

For the ALP search, one advantage of using the polarized light from
the galactic center is the large DM density.  In the case where a very
light scalar field (i.e., the ALP) plays the role of DM, we should
take into account the fact that the DM distribution cannot have
a non-trivial structure for the scale smaller than its de~Broglie
length.  Thus, at the galactic center, there exists a cut-off length
$\lambda$ below which the DM profile should become (almost)
flat.  The cut-off length $\lambda$ can be estimated by requiring that
the de~Broglie length for $r>\lambda$ be shorter than $r$; we estimate
$\lambda$ by solving $\lambda=1/(m_av_{\rm vir}(\lambda))$ with $v_{\rm
  vir} (r)$ being the virialized velocity at the position of $r$.  If
the BH mass dominates over the total DM mass within $r < \lambda$,
which is the case of our interest, the virial velocity is estimated as
\begin{align}
  v_{\rm vir} (r) = \sqrt{\frac{G M_{\mathrm{BH}}}{r}},
  \label{eq:virial}
\end{align}
where $G$ is the Newton constant and $M_{\mathrm{BH}}$ is the mass of
the BH, which results in
\begin{align}
  \lambda \sim 2.1 \times 10^2\, \mathrm{pc}
  \times \left( \frac{m_a}{10^{-21}\, \mathrm{eV}} \right)^{-2}.
\end{align}
Note that the source position $\sim 10^{-5}\,\mathrm{pc}$ is well
within the cut-off length for $m_a \lesssim 10^{-18}\, \mathrm{eV}$.
Then, the DM density at the position of the light source can be
evaluated by $\rho_{\rm DM}(r\sim\lambda)$, where $\rho_{\rm DM}$ is
the density profile of DM.  When we use the Einasto profile
\cite{Einasto:1965,Graham:2005xx}:
\begin{align}
  \rho_{\mathrm{Einasto}} (r) = \rho_s \exp \left[ \frac{-2}{\alpha}
  \left\{ \left( \frac{r}{r_s} \right)^\alpha - 1 \right\} \right],
\end{align}
with $\rho_s = 9 \times 10^6 M_\odot\ \mathrm{kpc}^{-3}$, $\alpha =
0.17$, and $r_s = 20\, \mathrm{kpc}$ \cite{Stadel:2008pn,
  Navarro:2008kc} for our galaxy, the ALP amplitude at the light
source is about $0$--$4$ orders of magnitude larger than that around
the earth for $m_a=10^{-22}$--$10^{-18}\ {\rm eV}$.  When we use the
Navarro-Frenk-White (NFW)
profile~\cite{Navarro:1995iw,Navarro:1996gj}:
\begin{align}
  \rho_{\mathrm{NFW}} (r) =
  \frac{\rho_H}{\frac{r}{R_H}\left( 1 + \frac{r}{R_H} \right)^2},
\end{align}
with $\rho_H=4 \times 10^7 M_\odot\ \mathrm{kpc}^{-3}$ and $R_H = 20\,
\mathrm{kpc}$ \cite{Klypin:2001xu}, the enhancement of the ALP
amplitude becomes more significant.  Such an enhancement of the ALP
amplitude is advantageous for the ALP search.

In using the polarized light from the galactic center (in particular,
from the region near Sgr A*), we should also take account of the
strong gravitational field at the source.  If the light source of our
concern is near the BH, the gravitational potential at the source
position should be sizable, resulting in a significant redshift of the
frequency of the ALP oscillation.  Besides, if the polarized light
is emitted from a region where the gravitational potential
significantly varies, the peak of $S(\omega, \delta)$ is broadened,
which makes ALP search more difficult.  Thus, although it may be
interesting to use the polarized light from Sgr A* for the ALP search,
more careful study is needed.  We leave it as a future task.

\subsection*{CMB foreground}

Now let us consider the possibility to use the data taken by CMB
polarization experiments, whose primary purpose is to find the
inflationary $B$-mode signals, to find ALPs.  We consider the
Fourier-space analysis of the time dependence of the polarization of
the CMB; for earlier discussion about the time dependence of the CMB
polarization due to the ALP DM, see \cite{Finelli:2008jv,
  Fedderke:2019ajk}.  We focus on the polarized foreground emission
below, although the polarized CMB (induced by the Thomson scattering)
may also be used as a probe of ALP as well.

It is well known that the CMB foreground emission from sources like
synchrotron radiation and Galactic dust is polarized.  The
polarization amplitude of such foreground emission is sizable and the
understandings of foregrounds are essential in the CMB experiments
searching for the inflationary $B$-mode signals.  The foreground
amplitude is strongly dependent on the CMB frequency, and many of the
CMB detectors have several frequency bands, some of which are (mainly)
dedicated to measuring the foreground amplitude.  Even though the
foreground emission is harmful for the $B$-mode detection, it may be
useful for the ALP search.  Hereafter, we estimate the sensitivity of
the CMB experiments to the signal of the ALP with the analysis using
the foreground polarization.

In the analysis, for simplicity, we assume that the CMB experiment
scans the sky region with the total solid angle of $\Omega$ within the
period of $\tau$.  The total solid angle $\Omega$ is divided into sky
patches each of which has a solid angle $\Delta\Omega$.  Then,
\begin{align}
  n_{\rm patch} = \frac{\Omega}{\Delta\Omega},
\end{align}
and the time available for scanning a single sky patch (within the
period of $\tau$) is
\begin{align}
  \Delta \tau \equiv \frac{\tau}{n_{\rm patch}}.
\end{align}
Then, using the relation between the polarization angle $\theta$ and
the Stokes parameters $Q$ and $U$, i.e., $\theta = \frac{1}{2}
\tan^{-1} (U/Q)$, $\sigma_N$, the uncertainty in the measurement
of the polarization angle, is estimated as
\begin{align}
  \sigma_N
  =
  \frac{{\rm NET}_{\rm arr}/\sqrt{\Delta\tau}}{\sqrt{2} P},
\end{align}
where ${\rm NET}_{\rm arr}$ is the noise-equivalent-temperature (NET)
of the detector array (which is given by ${\rm NET}_{\rm arr}={\rm
  NET}_{\rm det}/\sqrt{n_{\rm det}}$, with ${\rm NET}_{\rm det}$ and
$n_{\rm det}$ being the NET of single detector and the number of
detectors, respectively), and $P\equiv\sqrt{Q^2+U^2}$ is the
polarization amplitude~\cite{Akrami:2018mcd}.  Here, the uncertainties
of both $Q$ and $U$ are estimated to be $\sqrt{2}{\rm NET}_{\rm
  arr}/\sqrt{\Delta\tau}$, taking into account the factor of
$\sqrt{2}$ to convert the temperature noise to the polarization noise.
Although the polarization amplitude $P$ should depend on each sky
patch, for simplicity, we consider the case where the polarization
amplitudes for the patches of our interest are typically of the same
size. Then, the reach is estimated to be
\begin{align}
  g \gtrsim
  8.3 \times 10^{-14}\ {\rm GeV}^{-1}
  \times Z
  \left( \frac{m_a}{10^{-21}\ {\rm eV}} \right)
  \left( \frac{T}{1\ {\rm year}} \right)^{-1/2}
  \left( \frac{{\rm NET}_{\rm arr}}{10\ \mu{\rm K}\sqrt{\rm sec}} \right)
  \left( \frac{P}{10\ \mu{\rm K}} \right)^{-1}.
  \label{g_CMB}
\end{align}

There are on-going and future CMB experiments looking for the $B$-mode
signals with $T\sim$ a few years and ${\rm NET}_{\rm arr}$ of
$O(10)\ \mu{\rm K}\sqrt{\rm sec}$ or smaller.  The data available from
those experiments can be used for the ALP search.  The sensitivity of
each experiment depends on the details of the detector parameters, and
can be evaluated by using \eqref{g_CMB}.  For example, the $27\ {\rm
  GHz}$ detector of the small area telescope of Simons Observatory,
which will observe $10\ \%$ of the sky, will realize the angular
resolution of $25\ \mu{\rm K}\mbox{-}{\rm arcmin}$ with the
(effective) observation time of $1$ year \cite{Ade:2018sbj}.  Then,
${\rm NET}_{\rm arr}$ is estimated to be $36\ \mu{\rm K}\sqrt{\rm
  sec}$.  Using the RMS polarization amplitude of $P(30\ {\rm
  GHz})\simeq 7\ \mu{\rm K}$ \cite{Akrami:2018mcd}, Simons Observatory
may observe the ALP signal at $5\sigma$ level if $g \gtrsim 2.1\times
10^{-12}\ {\rm GeV}^{-1}\times (m_a/10^{-21}\ {\rm eV})$.  In more
future, CMB-S4 experiment may improve the sensitivity.  For the CMB-S4
experiment, the NET of single detector (for the frequency of $30\ {\rm
  GHz}$) is expected to be $177\ \mu{\rm K}\sqrt{\rm sec}$, while
$576$ detectors may become available for such a frequency
\cite{Abazajian:2019eic}, resulting in ${\rm NET}_{\rm arr}\simeq
7.4\ \mu{\rm K}\sqrt{\rm sec}$.  With $7$ years of observation, CMB-S4
may realize the $5\sigma$ discovery of the ALP signal if $g \gtrsim
1.7\times 10^{-13}\ {\rm GeV}^{-1}\times (m_a/10^{-21}\ {\rm eV})$.
The expected $5\sigma$ reach of CMB-S4 experiment after $7$-year
survey is shown in Fig.~\ref{fig:sensitivity} by the red line, taking
$\tau \sim 1\, \mathrm{day}$.  We comment here that the sensitivity
depends on the polarization amplitude $P$.  In particular, if the
observation is performed with a larger $P$, a better discovery reach
can be realized.

\section{Conclusions and Discussion}
\label{sec:conclusions}
\setcounter{equation}{0}

We have discussed the possibility to search for the signal of ALP DM
by observing the time dependence of the polarization of the light.  We
have proposed the Fourier-space analysis using the light from various
astrophysical sources.  We concentrated on the light from
astrophysical sources, but a similar analysis may be performed in the
proposed ALP search using optical
cavity~\cite{Obata:2018vvr,Zarei:2019sva} or gravitational wave
detector~\cite{Nagano:2019rbw}.  The discovery reach depends on the
performance of the detector (i.e., the time required to observe the
polarization of the light from a single sky patch and the size of the
noise in the observation) as well as the period available for the
observation.  If observed, the signal not only provides strong
evidence of ALP DM but also gives information about the ALP mass.

We listed several possible sources of polarized light from optical to
radio wave bands.  PPDs are good polarized sources in optical or
infrared band and it will give strong constraint on the axion coupling
constant if the long-time observation will be performed for $m_a\simeq
10^{-22}$--$10^{-19}$\,eV.  Some radio sources such as SNRs, radio
galaxies and pulsars are also possible candidates for such a purpose
and hence radio telescopes will also give comparable (or a bit weaker)
constraint if the observation time is long enough.  We also estimated
the reach of CMB polarization experiments, paying particular attention
to the possibility to use the foreground polarized emission.  The last
one is economical in the sense that the ALP search is possible
without affecting the operation of the mission to observe the
inflationary $B$-mode.

Our essential idea is to use time series data of the polarization
angle and find a peak in the Fourier space.  It has several advantages
as well as some cautions compared to that in the real time space.  In
the following, we list them in order:
\begin{itemize}

\item It may be possible to combine the results from different
  observations to improve the sensitivity because the position of the
  peak in the Fourier space is uniquely determined by the ALP mass and is
  independent of the observation (as far as the effect of the redshift
  is negligible).  In addition, if two (or more) experiments are done
  simultaneously, we may use the information about $\delta_*$ to check
  the consistency of the signal.

\item For the case where the source has a time dependence, the
  analysis in the Fourier space may help to discriminate the effects
  of the time dependence of the source from the signal.  We expect
  that the former does not mimic the latter unless the time dependence
  of the source is characterized by a very particular angular
  frequency with a width narrower than $\sim \pi/T$.  This is because,
  if the ALP DM exists, the signal function $S$ should have a very
  sharp peak at $\omega= m_a$ with the width of $\sim \pi/T$.

\item In our analysis, we assumed that the observation of the
  polarization angle is continuously performed during the observation
  period.  Such an assumption is just for simplicity and the
  qualitative behavior of the signal function $S$ holds in other
  cases.  The analysis using the signal function given in Eq.
  \eqref{fnS} applies to observations with unequal time steps.

\item Even if there is no ALP DM, the signal function $S$ fluctuates
  because of the noise, and $|S|$ has many local peaks.  The heights
  of some of the peaks may become accidentally much larger than the
  expected value, i.e., $\sqrt{n_{\rm tot}/2}$.  In deriving
  information about the ALP parameters from actual experiments, one
  should take into account look-elsewhere effects, whose study is
  beyond the scope of this letter.

\end{itemize}
In summary, Fourier-space analysis of the polarization of the light from
various sources may shed light on the properties of the ALP DM.

\vspace{2mm}
\noindent\underline{\it Acknowledgments:} The authors thank T. Fujita
and A. Kusaka for useful discussion.  This work was supported by JSPS
KAKENHI Grant (Nos.\ 17J00813 [SC], 16H06490 [TM], 18K03608 [TM],
18K03609 [KN], 15H05888 [KN] and 17H06359 [KN]).

\end{document}